\documentclass{article}

    \PassOptionsToPackage{numbers, compress}{natbib}

\usepackage[dblblindworkshop, final]{neurips_2026}
\workshoptitle{SaTQuML: Secure and Trustworthy Quantum Machine Learning}

\usepackage[utf8]{inputenc} 
\usepackage[T1]{fontenc}    
\usepackage{hyperref}       
\usepackage{url}            
\usepackage{booktabs}       
\usepackage{amsfonts}       
\usepackage{nicefrac}       
\usepackage{microtype}      
\usepackage{xcolor}         
\usepackage{amsmath}
\usepackage{braket} 
\usepackage{graphicx}
\usepackage{subcaption}
\title{Toward Joint Optimization of Circuit Depth and Training Data Size in Adaptively Grown Quantum Classifiers}

\begin{document}

\author{
  Saeefa Rubaiyet Nowmi \\
  Old Dominion University \\
  Norfolk, VA, USA \\
  \texttt{snowm001@odu.edu}
  \And
  Md Mahmuduzzaman Kamol \\
  Old Dominion University \\
  Norfolk, VA, USA \\
  \texttt{mkamo001@odu.edu}
  \And
  Mohammad Saidur Rahman \\
  University of Texas at El Paso \\
  El Paso, TX, USA \\
  \texttt{msrahman3@utep.edu}
}

\maketitle

\begin{abstract}
Building a quantum model involves a tradeoff: how complex the circuit should be, and how much training data it needs. Caro et al.\ show that models with fewer trainable gates need less training data to generalize well. Q-FLAIR shows that a quantum feature-map circuit can be grown gate-by-gate, stopping once further growth stops improving the training loss. We ask whether these two results combine into a predictable scaling law. Does Q-FLAIR's own stopping rule pick larger or smaller circuits as training data grows? Does the resulting generalization behavior track Caro et al.'s bound?

We reimplement Q-FLAIR's growth mechanism faithfully, including its analytic reconstruction and exact stopping rule. We run it on full-resolution (784-pixel) MNIST 3-vs-5 classification, at five training-set sizes from $N=2000$ to $10000$. We then fine-tune each resulting circuit, so we can measure Caro et al.'s notion of active gates, $K$.

We find no predictable relationship between training-set size and the circuit size Q-FLAIR converges to. Circuit size and test accuracy both vary non monotonically with $N$, and seed-to-seed variance is nearly as large as any trend across $N$. The empirical generalization gap never exceeds Caro et al.'s bound in 14 of 15 runs, so the bound holds as a valid guarantee. But the gap correlates only weakly with the bound's value ($r=0.12$). This shows that $K$ does not explain most of the variation we observe. Why a valid guarantee can coexist with such weak predictive power remains an open question, and answering it may be necessary before circuit depth and training data size can be jointly optimized in practice.
\end{abstract}

\section{Introduction}

Quantum feature maps embed classical data into quantum states, and this idea underlies most modern quantum machine learning \cite{havlivcek2019supervised, schuld2019quantum}. A common design choice re-encodes data across multiple layers, a technique known as data re-uploading \cite{perezsalinas2020data}. Building such a model requires two largely independent decisions. The first is how much training data to use. The second is how complex the variational circuit should be.

Caro et al.~\cite{caro2022generalization} give a theoretical answer to the second question, given the first. A model with $T$ trainable gates generalizes with error $O(\sqrt{T\log T/N})$. This tightens to $O(\sqrt{K\log(MT)/N})$ when only $K\ll T$ gates change substantially during training. Q-FLAIR~\cite{jager2026qflair} answers a complementary question. Given a fixed training set, it grows a feature-map circuit gate-by-gate from empty. It uses an efficient analytic reconstruction to pick each gate's best weight and data feature, and it does this classically. Growth stops once further growth no longer improves the training loss.

Adaptive circuit growth is not unique to Q-FLAIR. ADAPT-VQE grows ansatze gate-by-gate for molecular simulation \cite{grimsley2019adaptive}, and VAns grows and prunes circuit structure during training \cite{bilkis2021semiagnostic}. Caro et al.\ use VAns in their own unitary compiling experiments. Neither ADAPT-VQE nor VAns, however, is studied as a function of training-set size, and neither work asks the question their combination with a generalization bound raises.

That question is this: if Q-FLAIR's growth algorithm is handed different amounts of training data, does the circuit size it converges to change in a way consistent with Caro et al.'s theory? A related question also follows: is the number of trainable gates required systematically correlated with the size of the training data? Answering these questions requires actually running Q-FLAIR's mechanism, not a fixed circuit family, across a range of $N$. This is what we do here, on full-resolution (unreduced, 784-pixel) MNIST 3-vs-5 classification. This setting matches Q-FLAIR's own ``MNIST 28$\times$28'' benchmark dimensionality exactly \cite{jager2026qflair}.

\textbf{Contributions.}
\textbf{(1)} A faithful reimplementation of Q-FLAIR's QNN growth mechanism: the exact gate pool, observable, reconstruction, and stopping rule. We make this tractable at $d=784$ by using a bounded scalar optimizer in place of brute-force grid search for the classical feature-selection step.\\
\textbf{(2)} A joint fine-tuning stage that makes Caro et al.'s $K$ measurable for a Q-FLAIR-grown circuit, something neither source method provides on its own. It includes an explicit safety comparison against Q-FLAIR's own pre-fine-tune result.\\
\textbf{(3)} An empirical finding: on high-dimensional raw-pixel data, circuit size and accuracy vary non-monotonically and noisily with $N$, with no discernible scaling trend. This indicates that a straightforward joint depth/data scaling relationship does not emerge automatically from combining these two methods, and that achieving one in practice will require further exploration.
\section{Related Work}

Caro et al.~\citep{caro2022generalization} validate their generalization
bound through two numerical studies: a quantum convolutional neural
network applied to phase classification, and a separately-grown circuit
(VAns) applied to unitary compiling. In both cases, the circuit
architecture is either fixed a priori or grown once and subsequently
held constant, with the training-set size $N$ never varied for a given
architecture. Neither study therefore examines how an adaptive growth
algorithm's own output would respond to changes in the amount of
available training data.

Q-FLAIR~\citep{jager2026qflair} grows a feature-map circuit gate-by-gate
from an empty ansatz. At each iteration, it employs an analytic
reconstruction to select the best-performing gate, weight, and data
feature entirely through classical computation, terminating once further
growth ceases to improve the training loss. Across their reported
experiments, $N$ is held fixed for each task under consideration, and
the question of how the algorithm's converged output depends on $N$ is
not addressed.

The present work addresses this gap directly. We execute Q-FLAIR's
growth mechanism, without modification, across a range of values of $N$,
and characterize how its output varies as a result. We further introduce
a short post-hoc fine-tuning procedure, which renders Caro et al.'s
notion of active gates, $K$, measurable for the circuits that Q-FLAIR
produces.
\section{Method}

\textbf{Growth (Phase 1).} We reimplement Q-FLAIR's gate pool exactly (their Eq.~25): $V=\{R_x(\theta), R_y(\theta), R_x(\theta,x_k), R_y(\theta,x_k), R_{xx}(\theta), R_{yy}(\theta), R_{xx}(\theta,x_k), R_{yy}(\theta,x_k), H\}$. We place these gates on a 5-qubit ring topology. This matches Q-FLAIR's own QNN qubit ceiling. 

At each iteration, we consider every candidate gate. For each one, we reconstruct its model output as a closed-form sinusoid of its rotation angle. This takes only 2 additional quantum evaluations. We then solve for the loss-minimizing weight classically.

At $d=784$ candidate features, a brute-force grid search over feature index is too slow. So we use a bounded scalar minimizer, with convergence tolerance $10^{-3}$. This cuts the classical cost by roughly an order of magnitude, solving the same one-dimensional bounded optimization problem that Q-FLAIR specifies.
Growth stops the first time the best candidate's training-loss improvement falls below $\Delta L=10^{-3}$. We place no additional cap on circuit size.

\textbf{Joint fine-tune (Phase 2).} Q-FLAIR freezes each gate's weight
once found, so Caro et al.'s notion of gates changing during
optimization does not apply to Phase 1 alone. To measure it, we
warm-start from the finished circuit and run a joint Adam fine-tune over
all weights, using mini-batches and early stopping, defining $K$ as the
number of gates whose weight moves by more than $\delta=0.05$.

We record the validation loss at the original Phase-1 weights as a
baseline before fine-tuning, and afterward keep whichever weights,
original or fine-tuned, give the lower validation loss; Phase 2 can
therefore never report a worse model than Phase 1 alone.

We report the full Caro et al.\ Theorem 3 bound:
$\min_{K}[\sqrt{K\log(MT)/N} + \sum_{k>K}M\Delta_k]$. When fine-tuning
makes no improvement, every $\Delta_k$ is exactly zero and the bound
would trivially equal zero, which does not reflect a real
generalization guarantee; we flag such cases as not applicable.
\section{Experimental Setup}

We use MNIST digits 3 vs.\ 5 \citep{jager2026qflair}: raw, unreduced
784-pixel intensities, rescaled to $[-\pi,\pi]$ using training-set
statistics. We sweep $N\in\{2000,4000,6000,8000,10000\}$ with 3 seeds
per $N$, yielding 15 runs, each evaluated on the same held-out test set
of 1000 examples.

Our loss is the negative log-likelihood, matching both source papers'
convention; the decision threshold is set via a classical ROC-style
search maximizing balanced accuracy on the training set, as in Q-FLAIR.
All circuits are simulated exactly; each run takes minutes to tens of
minutes, depending on $N$ and gate count.

We implement circuit construction, statevector simulation, and the
Phase-2 Adam fine-tune in PennyLane; Phase 1's classical feature-weight
search uses SciPy's bounded scalar minimizer.
\section{Results}

\begin{table}[t]
\caption{Per-$N$ summary across 3 seeds, full-resolution (784-pixel) MNIST.}
\label{tab:summary}
\centering
\begin{tabular}{rccc}
\toprule
$N$ & $T^*$ (mean $\pm$ SD) & test acc.\ (mean $\pm$ SD) & gen.\ gap (mean $\pm$ SD) \\
\midrule
2000  & 22.67 $\pm$ 9.07 & 0.892 $\pm$ 0.028 & 0.0089 $\pm$ 0.0047 \\
4000  & 16.33 $\pm$ 2.52 & 0.841 $\pm$ 0.008 & 0.0141 $\pm$ 0.0123 \\
6000  & 20.67 $\pm$ 4.73 & 0.892 $\pm$ 0.027 & 0.0045 $\pm$ 0.0028 \\
8000  & 18.00 $\pm$ 1.00 & 0.882 $\pm$ 0.001 & 0.0048 $\pm$ 0.0024 \\
10000 & 22.67 $\pm$ 8.96 & 0.874 $\pm$ 0.022 & 0.0128 $\pm$ 0.0101 \\
\bottomrule
\end{tabular}
\end{table}

\subsection{No discernible scaling trend}

\begin{figure}[t]
    \centering

    \begin{subfigure}[t]{0.24\textwidth}
        \centering
        \includegraphics[width=\linewidth]{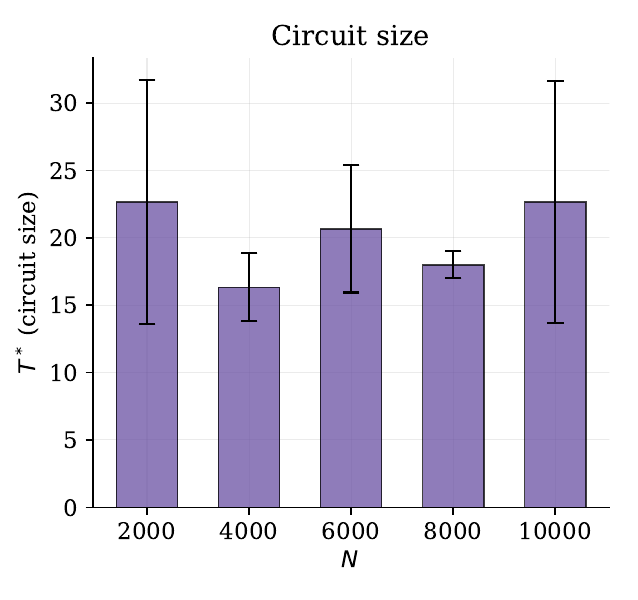}
        \caption{}
        \label{fig:circuit_size}
    \end{subfigure}
    \hfill
    \begin{subfigure}[t]{0.24\textwidth}
        \centering
        \includegraphics[width=\linewidth]{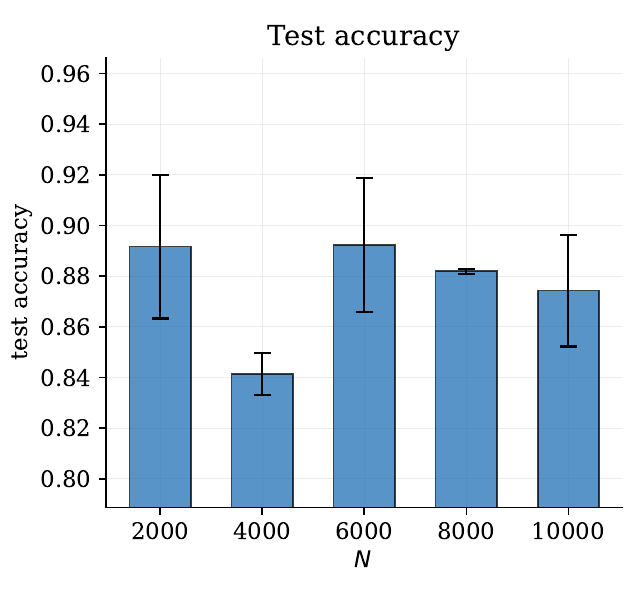}
        \caption{}
        \label{fig:test_accuracy}
    \end{subfigure}
    \hfill
    \begin{subfigure}[t]{0.24\textwidth}
        \centering
        \includegraphics[width=\linewidth]{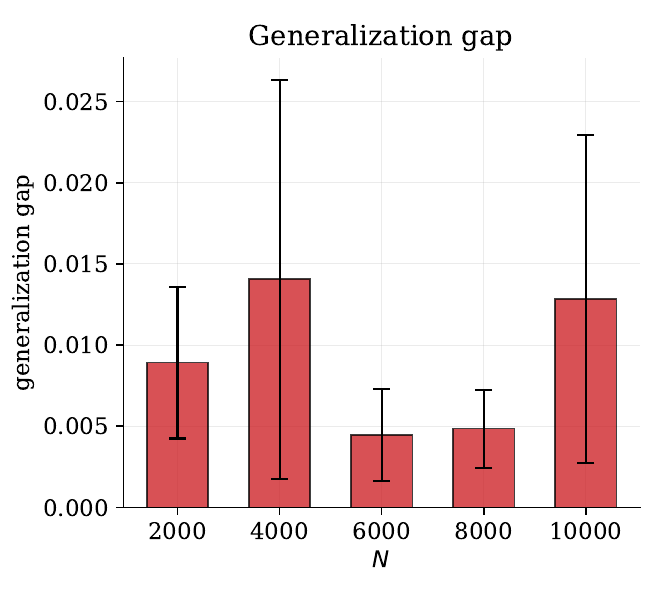}
        \caption{}
        \label{fig:generalization_gap}
    \end{subfigure}
    \hfill
    \begin{subfigure}[t]{0.24\textwidth}
        \centering
        \includegraphics[width=\linewidth]{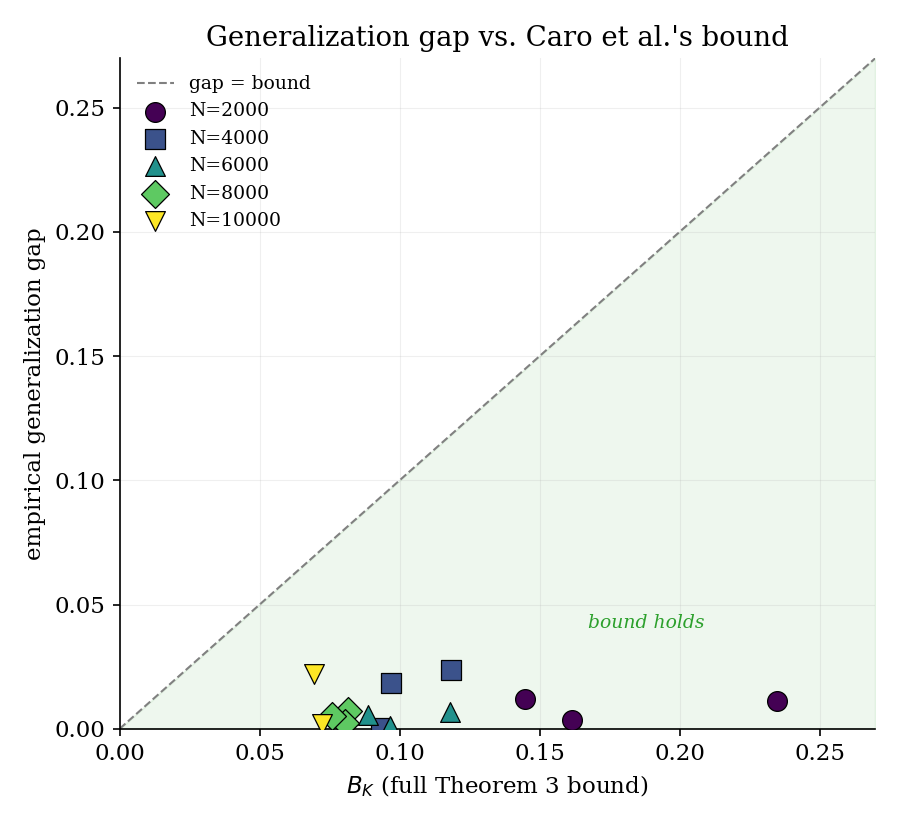}
        \caption{}
        \label{fig:caro}
    \end{subfigure}

    \caption{
    Relationship between training-set size $N$ Vs
    (a) circuit size $T^*$,
    (b) test accuracy,
    (c) generalization gap, and
    (d) Generalization gap versus the full Caro et al.\ Theorem~3 bound $B_K$.
    The shaded region indicates where the bound holds.
    }
    \label{fig:combined_four}
\end{figure}


We might expect more data to support a larger, more accurate circuit. Our results do not support this expectation. Table~\ref{tab:summary} and Figure~\ref{fig:combined_four}\subref{fig:circuit_size}--\subref{fig:generalization_gap} show neither $T^*$ nor test accuracy varies monotonically with $N$.

Circuit size ranges between 16 and 23 gates, with no clear pattern across $N$. Accuracy ranges between 84\% and 89\%, also with no clear pattern. The seed-to-seed standard deviation is as large as $\pm9$ gates. This is comparable to the total spread we see across all five values of $N$.

\subsection{Consistency with Caro et al.'s bound}

One run out of 15 ($N{=}10000$) made no improvement during fine-tuning. All its gate weights stayed identical before and after, so $K=0$ by construction. For this run, the bound would trivially equal zero. We exclude it from this analysis. That leaves 14 applicable runs. 

Across these, the empirical gap never exceeds the full bound (Figure~\ref{fig:caro}). But the two are only weakly correlated: $r=0.12$. This is not better than using raw $T$ alone ($r=0.05$). So the bound holds as valid. But it is not a useful predictor of the actual gap, at this scale and dimensionality.

\section{Discussion}

We attribute the absence of an $N$-dependent trend to the dimensionality of the candidate feature space: with only ten candidate features, a fixed, absolute training-loss-improvement threshold would interact visibly with $N$, since a noisy loss estimate at small $N$ would let marginal candidates clear the threshold by chance and inflate circuit size, whereas our actual pool of 784 raw-pixel candidates is large enough that some gate clears the same threshold at almost any $N$, decoupling the stopping decision from data availability. This suggests that combining an adaptive growth algorithm with a data-efficiency motivation does not, by itself, guarantee a joint scaling relationship; feature-space dimensionality instead acts as a confounding factor that must be addressed explicitly.\\
A more principled stopping rule might scale its threshold with both $N$ and the
candidate pool size, for instance by correcting for the number of
comparisons made per iteration, a correction that may be necessary
before a genuine joint depth-data scaling law can emerge on
high-dimensional inputs.
\section{Limitations}

Our results come from a single dataset and task. We do not know whether these observations generalize to other high-dimensional inputs, or to
Q-FLAIR's QSVM variant. We use three seeds per $N$, and this limits our statistical power. A trend smaller than the observed seed variance could exist, and we would not detect it. Phase 2 is an addition from this work; Q-FLAIR itself does not include a fine-tuning stage. We chose its learning rate and its active-gate threshold $\delta$ once, and we did not test other values. Different choices could change the measured value of $K$, and this could in turn affect our reported correlation with Caro et al.'s bound.

\section{Conclusion}

This work examined whether Q-FLAIR's gate-by-gate growth mechanism, executed at five training-set sizes on full-resolution MNIST 3-vs-5 classification, yields circuit sizes or generalization behavior consistent with a joint depth-data scaling relationship motivated by Caro et al.'s theory. No such relationship was observed: both circuit size and accuracy vary non-monotonically with $N$, and although Caro et al.'s bound is never violated in 14 of 15 runs, it correlates only weakly with the observed gap ($r=0.12$), an open question in its own right. Joint optimization of circuit depth and training data size cannot, therefore, be assumed to emerge automatically from an adaptive growth algorithm and a generalization bound alone; the dimensionality of the candidate feature pool is a plausible contributing factor warranting further investigation.

\bibliographystyle{plainnat}
\bibliography{reference}



\end{document}